\documentclass[12pt]{article}
\usepackage{eurosym,geometry,ulem,graphicx,color,setspace,comment,footmisc,natbib,array,hyperref}
\usepackage{amssymb,amsbsy,amsmath,amsfonts,amssymb,amscd} 
\usepackage[T1]{fontenc}  
\usepackage{booktabs} 
\usepackage{tabularx} 
\usepackage[font=small,labelfont=bf,tableposition=top]{caption} 
\usepackage{subcaption} 
\usepackage{multirow} 
\usepackage{caption} 

\usepackage{adjustbox} 
\usepackage{makecell} 

\usepackage{float} 
\usepackage{pdflscape} 
\usepackage{natbib} 
\usepackage{bbm} 
\usepackage{lscape} 

\usepackage{url}
\usepackage{multirow}
\usepackage{footnote}
\usepackage{framed}
\usepackage{enumerate}
\usepackage{stmaryrd}
\usepackage[toc,page]{appendix}
\usepackage{dsfont}
\usepackage{threeparttable}
\usepackage{epstopdf}
\usepackage{enumitem}
\usepackage{changepage}
\usepackage{indentfirst}

\newcolumntype{L}[1]{>{\raggedright\let\newline\\arraybackslash\hspace{0pt}}m{#1}}
\newcolumntype{C}[1]{>{\centering\let\newline\\arraybackslash\hspace{0pt}}m{#1}}
\newcolumntype{R}[1]{>{\raggedleft\let\newline\\arraybackslash\hspace{0pt}}m{#1}}

\providecommand{\keywords}[0]
{
  \small	
  \noindent 
  \textbf{Keywords:}
}
\providecommand{\JELcodes}[0]
{
  \small	
  \textbf{JEL classification:}
}

\title{The growth effects of tropical cyclones in the U.S.: new evidence from state to county level}
\author{Eric Kulanthaivelu}
\date{August 2023}
\author{Eric Kulanthaivelu\thanks{Centre d\textquoteright Economie et de Management de l\textquoteright Oc\'{e}an Indien, Universit\'{e} de La Reunion, 15 avenue Ren\'{e} Cassin, 97715, Saint-Denis Cedex 9, France \newline  
\textit{E-mail: eric-sivatharsan.kulanthaivelu@univ-reunion.fr}}}

\begin{document}
\maketitle
\begin{abstract}

\noindent Tropical cyclones have always been a concern for public authorities in the U.S., with a season lasting nearly half of the year. Using longitudinal data on economic growth and exposure to tropical cyclones, we provide new comprehensive analyses of these disasters' growth impact in U.S. states and counties. While we find that results remain insignificant with country-wide samples, this study shows that the effects can be, instead, significant at regional level. In fact, a specific analysis of Florida State stresses significant short-term negative effects, both at county and state levels. At this stage, this study calls for an in-depth investigation of the underlying mechanisms leading to such growth depletion but already pins down the important heterogeneity in the effects of a catastrophe on growth patterns within a country.

\end{abstract}

\keywords Natural disasters; Cyclones; Growth; Economic impacts; Environmental issues \\
\JELcodes O44; Q54; R11

\newpage

\section{Introduction}

Natural disasters commonly designate catastrophic events caused by natural processes of the Earth. These phenomena are often distinguished between those of geological, climatic or even biological nature. Throughout history, their consequences on human systems and societies have always been felt, and the ability to predict or prepare for them has accordingly represented a perennial challenge. In the context of assessing the impacts of natural disasters, the U.S. turns out to be an interesting case study as it is exposed to a wide range of disasters such as volcanic eruptions, earthquakes, winter storms, tornadoes, wildfires, droughts etc. This paper particularly focuses on tropical cyclones, which are arguably the costliest and the most threatening for material assets. 

Among other countries, the U.S. regularly vied for the attention of researchers in the empirical literature. Some single-event studies such as those on the 2005 Hurricane Katrina examine the impact on jobs or population displacement, housing market, education or health outcomes (Vidgor, 2008; Groen \& Polivka, 2008; Sacerdote, 2012; Deryugina \& al., 2018; Deryugina \& Molitor, 2020). More closely related to the present study, Strobl (2011) evaluates the impact of hurricanes on U.S. coastal areas' economic growth from 1970 to 2005. On the year the hurricane occurs, a detrimental effect is estimated at county level, while this negative impact is only detected on the quarter of occurrence and is offset by the end of the year at state level. Finally, no significant effect is perceived at national level.

Using annual and quarterly panel data sets of U.S. counties' or states' economic growth respectively from 1970 to 2020, this paper provides a reappraisal of the latter research question in the context of tropical cyclones for different levels of exposure. To achieve this, we exploit a geophysical data set to build a physical intensity measurement for all tropical cyclones that affect the U.S. at some point in their trajectory. Then, as tropical cyclones not only affect directly impacted areas but also neighboring territories, we resort to spatial panel models as in Strobl (2011) to take into consideration the influence of contiguous counties' or states' growth on each area under study. Overall, across our different panels of states or counties, we find that tropical cyclones do not disrupt economic activity. In a second phase of the study, we focus on one particular state, namely Florida. The interest in this State is not new in the empirical literature. Belasen \& Polachek (2009) stress the negative impact on local employment in the wake of hurricanes. This detrimental effect leads, in counterpart, to a positive relationship between hurricane occurrence and earnings in the short-run. Brown \& al. (2021) evaluate revenue losses within the tourism sector due to tropical cyclones and highlight the increased effect on waterfront counties compared to those located deeper in Floridian lands. More recently, Pollack \& Kauffmann (2022) demonstrate how cyclonic risk is imperfectly captured in Florida Keys' housing prices. Unlike other states in the U.S., Florida combines several characteristics that might amplify tropical cyclones' impacts. First, it is the most frequently exposed state in the U.S., while it is one of the richest states and one of the most populated. The latter combination of factors is even more concerning than Florida's largest metropolitan areas such as Miami are all located close to the shoreline where tropical cyclones make landfall with the greatest intensity. Finally, as pointed out in the above literature review, Florida has an increased vulnerability due to the importance of tourism for its economy. This single-state assessment is carried out by means of time series analysis which constitutes one of the novelties in this paper. After having a built stationary measurement for quarterly economic growth, our modeling concludes with a continuous negative effect over a year, which is merely significant on the quarter the cyclone occurs and two quarters later. These conclusions at state level are emphasised by complementary results on Floridian counties, which suggest that growth penalties accumulate over three years on a smaller scale. At last, the main contributions of this paper originate from its complete replicability, the use of updated data up to 2020, a twofold econometric strategy and a specific focus on Florida.

The paper is structured as follows. Section~\ref{sec2} introduces the cyclone and economic data, and describes construction of a stationary measure of economic growth at state level. Section~\ref{sec3} presents the methodology as well as the main results on the effects of tropical cyclones on U.S. economic growth at different levels while Section~\ref{sec4} examines the specific case of Florida. Finally, Section~\ref{sec5} concludes.

\section{Data}
\label{sec2}
This section introduces the data. First, we present those exploited in order to build our tropical cyclone intensity measurement. Then, the set of economic data at the state and county levels is presented. As the analysis of personal income per capita data at state level discloses the existence of seasonality, all the transformations made to address this issue are described step-by-step. 

\subsection{Cyclone data}
\begin{figure}[H]
    \centering
    \includegraphics[width = 1\textwidth]{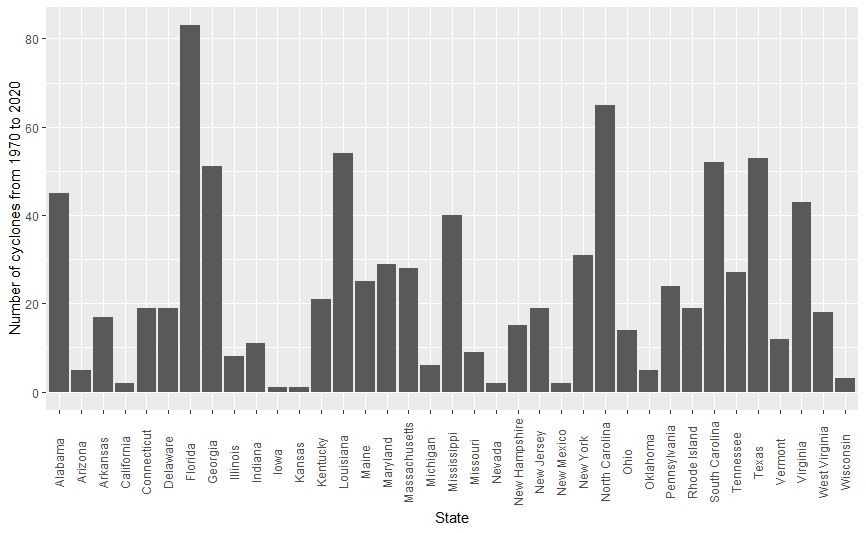}
    \caption{Total number of cyclone events from 1970 to 2020 in the contiguous U.S., by state.}
    \label{fig1}
\end{figure}

Our primary cyclone data come from from Tropical Cyclone Exposure Database (TCE-DAT; Geiger \& al., 2018), with an extension up to 2020 kindly provided by the research team upon request. TCE-DAT is a longitudinal data set on each landfalling tropical cyclone event worldwide from 1950 to 2020. As this paper focuses on the U.S., we only extract data related to this country, \textit{i.e.} the events that have affected the U.S. at some point of their trajectory. TCE-DAT reports wind speed and location data for each cyclone event that has reached at least 34 knots ($\approx$ 63 km/h) wind speed for at least one minute. It actually relies on the set of events reported by the widely used International Best Track Archive for Climate Stewardship (IBTrACS; Knapp \& al., 2010), maintained by the National Climatic Data Center of the National Oceanic and Atmospheric Administration (NOAA). Using the 6-hourly observations on latitude and longitude position, wind speed, surface air pressure and trajectory of cyclone centres (called the "eye") provided by the IBTrACS, TCE-DAT builds overall cyclone trajectories and intensities at 0.1° latitude x 0.1° longitude grid cell level for each of the aforementioned event by applying Holland's (2008) wind field model. Hence, we consider here all the events during which at least one grid cell of the U.S. is hit according to the simulated wind field. Due to economic data limitations, our panel starts in 1970 and ends in 2020. Florida is the most frequently hit State during this time period with 83 events recorded, followed by North Carolina, Louisiana and Texas, with respectively 65, 54 and 53 cyclone events (Figure~\ref{fig1}).

\begin{figure}[H]
    \centering
    \includegraphics[width = 0.8\textwidth]{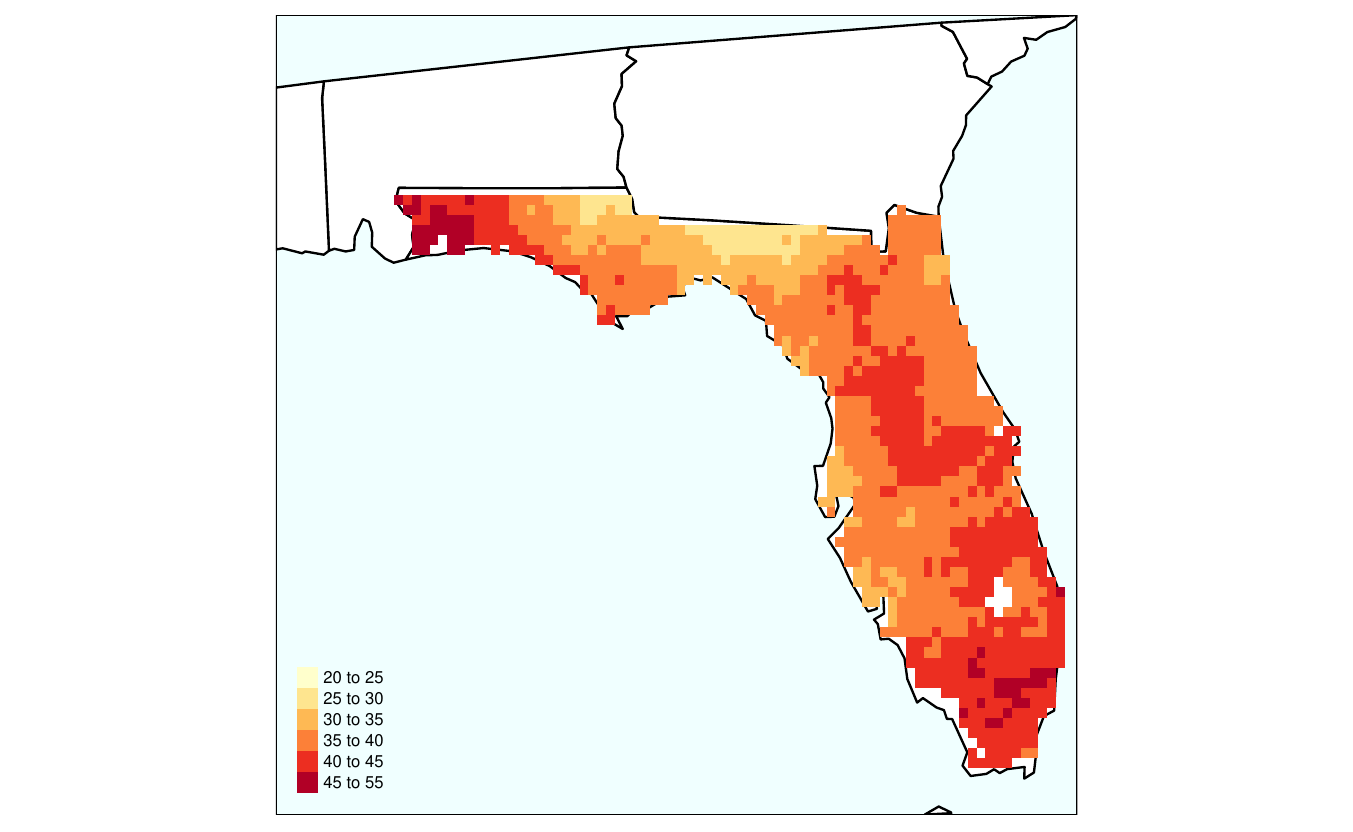}
    \caption{Yearly average wind speed at pixel level in Florida from 1970 to 2020.}
    \label{fig2}
\end{figure}

Apart from being the most exposed State in the U.S., Florida is also characterized by a wide range of factors of vulnerability regarding cyclonic risk. These factors are detailed, in-depth, in section~\ref{sec4} and constitute the reason why we choose to focus on this specific state in a second part of the study. Among others, the exposure to cyclonic risk is homogeneous in Florida, as it is shown in Figure~\ref{fig2}. The yearly average wind speed over Florida calculated from 1970 to 2020 shows small dispersion as this State is often entirely affected when it is hit by a tropical cyclone, unlike Texas, for instance, where all areas are not exposed to cyclonic risk.

\subsection{Economic data}

Data on personal income are taken from the Bureau of Economic Analysis’s (BEA) regional economic accounts. More specifically, we extract data on Quarterly Personal Income by State as well as Local Area Personal per Capita Income, which provides annual county-level estimates from 1969 to 2020. The BEA defines personal income as the income received by people from labor, properties, financial income such as interests or dividends, rents, and government benefits. As personal income data are released in current US\$ values, these have been converted to constant 2020 US\$ using the consumer price index reported by the U.S. Bureau of Labor Statistics. State level personal income per capita data are obtained after including state population data from the U.S. Bureau Census (2021) to our data set. 

Having said that, as we are in the context of panel data, it is necessary to assess whether these personal income per capita series are stationary. To do this, we first run panel and individual unit root tests for each series (logged) in levels and in first-difference for State level quarterly personal income per capita data. Table~\ref{tab1} outlines the results. According to the individual unit root tests, \textit{i.e.} the Augmented Dickey-Fuller (ADF) and Philips-Perron (PP) test, and one of the panel unit root tests, \textit{i.e.} the Im-Pesaran-Shin (IPS) test, first differencing the series seems to remove the unit root. However, the Levin-Lin-Chu (LLC) panel unit root test suggests that there is a problem as the null hypothesis of having at least one non stationary series among the panel is accepted with a P-value of 1. 
\begin{table}[H]
    \centering
    \caption{Unit root rests for raw series}
    \label{tab1}

    \begin{adjustbox}{width=1\textwidth}
        \begin{tabular}{l c|l c} 
        \hline \hline
            \underline{\textit{Individual unit root tests}} & Fraction of states for which test & \underline{\textit{Panel unit root tests}} & P-value of the test\\
             & is rejected at the 10\% level & & \\
            \hline
            1) Personal income per capita (logged) in levels & & 1) Personal income per capita (logged) in levels & \\
            \textit{Augmented Dickey-Fuller test} & 0/37  & \textit{Levin-Lin-Chu test} & 0.4705 \\
            \textit{Philips-Perron test} &  2/37 & \textit{Im–Pesaran–Shin test} & 1.0000 \\
            & & \\
            2) Personal income per capita (logged) in first-difference & & 2) Personal income per capita (logged) in first-difference & \\
            \textit{Augmented Dickey-Fuller test} & 37/37 & \textit{Levin-Lin-Chu test} & \textbf{1.0000} \\
            \textit{Philips-Perron test} & 37/37 & \textit{Im–Pesaran–Shin test} & 0.0000 \\
        \hline \hline 
        \end{tabular}
        
    \end{adjustbox}
\footnotesize \textit{Notes: }Augmented Dickey-Fuller tests with 5 lags and trend. Philips-Perron tests with 5 lags and trend. \\ 
Levin-Lin-Chu tests with 3 lags and trend, $H_0 =$ panels contain unit roots \\
Im–Pesaran–Shin tests with 3 lags and trend, $H_0 =$ all panels contain unit roots
\end{table}

\begin{figure}[H]
    \centering
    \includegraphics[width = 0.65 \textwidth]{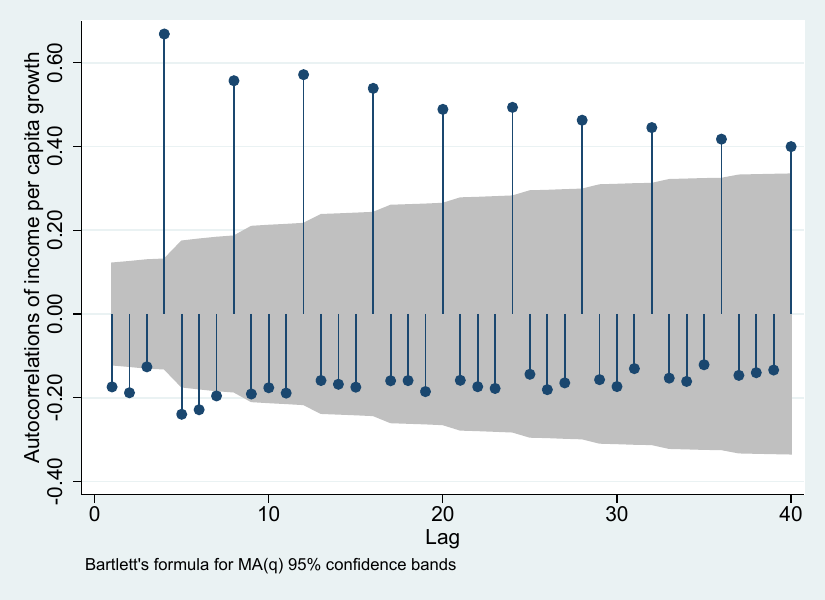}
    \caption{ACF for Floridian quarterly personal income per capita growth rate}
    \label{fig3}
\end{figure}

To deal with this issue, we compute the auto-correlation function (ACF) of one of our series and analyse the results. Figure~\ref{fig3} plots the ACF obtained for Florida, and the latter highlights the presence of seasonality, which explains the results presented in Table~\ref{tab1}. 

\begin{figure}[H]
    \centering
    \includegraphics[width = 0.49 \textwidth]{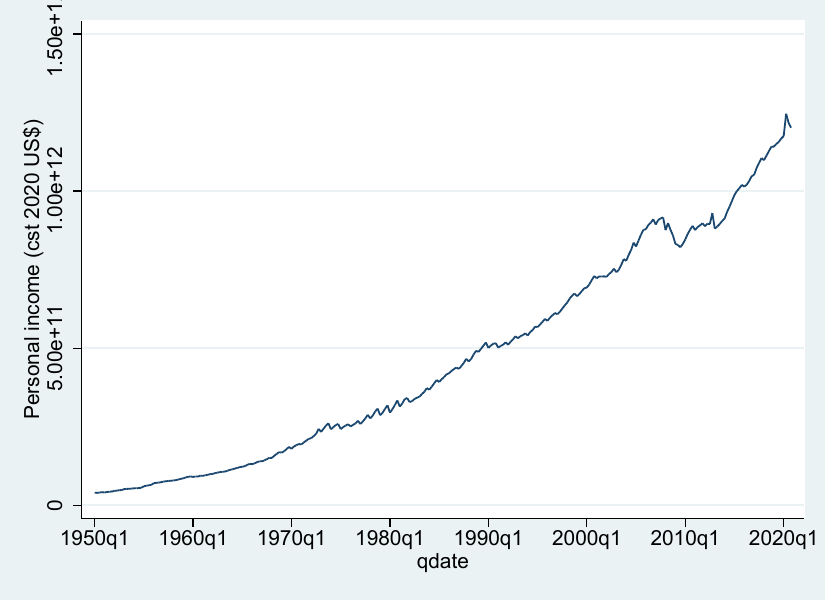}
    \includegraphics[width = 0.49 \textwidth]{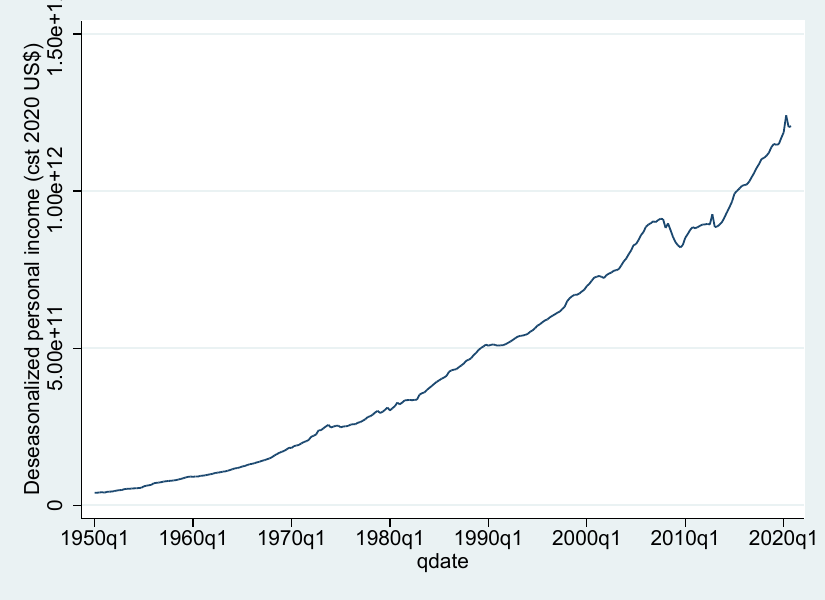}
    \caption{Plots for Floridian quarterly personal income through time. Deseasonalized series on right hand-side.}
    \label{fig4}
\end{figure}

Finally, to remove the seasonality in the data, we use the X-13-ARIMA-SEATS program developed jointly by the U.S. Census Bureau and the Bank of Spain. The result of this seasonal adjustment program is shown in Figure~\ref{fig4} in the case of Florida, and Table~\ref{tab2} now shows that first-differences of deseasonalized series are all stationary.

Regarding our cyclone data, it is found to be $I(0)$ across all unit root tests in levels, and further examination of its ACF shows that this series is arguably exogenous. As for county level annual personal income per capita, first differencing the series is enough to render them stationary.\footnote{All the statistical tests related to both of these variables are available upon request.}

\begin{table}[H]
    \centering
     \caption{Unit root rests for deseasonalized series}
\label{tab2}

    \begin{adjustbox}{width=1\textwidth}
        \begin{tabular}{l c|l c} 
        \hline \hline
             \underline{\textit{Individual unit root tests}} & Fraction of states for which test & \underline{\textit{Panel unit root tests}} & P-value of the test\\
             & is rejected at the 10\% level & & \\
            \hline
            1) Personal income per capita (logged) in levels & & 1) Personal income per capita (logged) in levels & \\
            \textit{Augmented Dickey-Fuller test} & 0/37  & \textit{Levin-Lin-Chu test} & 0.0001 \\
            \textit{Philips-Perron test} &  0/37 & \textit{Im–Pesaran–Shin test} & 0.9972 \\
            & & \\
            2) Personal income per capita (logged) in first-difference & & 2) Personal income per capita (logged) in first-difference & \\
            \textit{Augmented Dickey-Fuller test} & 37/37 & \textit{Levin-Lin-Chu test} & \textbf{0.0000} \\
            \textit{Philips-Perron test} & 37/37 & \textit{Im–Pesaran–Shin test} & 0.0000 \\
        \hline \hline 
        \end{tabular}
    \end{adjustbox}
\footnotesize \textit{Notes: }Augmented Dickey-Fuller tests with 5 lags and trend. Philips-Perron tests with 5 lags and trend. \\ 
Levin-Lin-Chu tests with 3 lags and trend, $H_0 =$ panels contain unit roots \\
Im–Pesaran–Shin tests with 3 lags and trend, $H_0 =$ all panels contain unit roots
    
\end{table}

\section{Measuring the impact of tropical cyclones on economic growth from state to county level}
\label{sec3}
This third section outlines our methodology and main results. After having described the construction of our cyclone intensity predictor, this section presents the econometric model selected to measure the impact of tropical cyclones on economic activity and the obtained results. In a last subsection, we check the robustness of our main results for a variety of changes in the empirical framework.

\subsection{Cyclone intensity measurement}
\label{sec3.1}
This paper relies on a physical measure of tropical cyclones that is measured quarterly to fit with our state level economic data time interval, and annually for our analyses at county level. It takes the form of a population-weighted average of the maximum sustained wind speed over all pixels in each state - respectively, county - in a given sample. Mathematically, one can write the indicator at state level as:

\begin{equation}
    \overline{Cyc}_{i,q} = \dfrac{\sum \limits_{p=1}^{P_i} \max\limits_{s \in \{1,...,S_{p,i,q}\}}\{Wind_{p,s,i,q}\}Exposed~Population_{p,q}}{Total~Population_{i,q}}
    \label{eq1}
\end{equation}
where $Wind_{p,s,i,q}$ is the wind speed recorded in pixel $p$, for storm $s$ that struck state $i$ in quarter~$q$. $Exposed~Population_{p,q}$ corresponds to the estimated population in pixel $p$ and reported in TCE-DAT, while $Total~Population_{i,q}$ is the number of inhabitants in state $i$ during quarter $q$ as estimated in the U.S. Bureau Census (2021) data set. All pixel values are then averaged across the total number of pixels per state $P_i$.

As for the yearly indicator, it includes a temporal weight $(12-m_{p,s})/12$ as in Noy (2009), this additional parameter is incorporated to capture the contemporaneous impact with more accuracy as a cyclone that strikes in the early months of a given year has arguably a bigger impact in the same year than another one - with same intensity - occurring some months later. This augmented version of equation~(\ref{eq1}) can be written as follows:

\begin{equation}
    \overline{Cyc}_{c,t} = \dfrac{\sum \limits_{p=1}^{P_c} \max\limits_{s \in \{1,...,S_{p,c,t}\}}\{Wind_{p,s,c,t}\}\left(\dfrac{12-m_{p,s}}{12}\right)Exposed~Population_{p,t}}{Total~Population_{c,t}}
    \label{eq2}
\end{equation}
with $m_{p,s}$ the month in which the storm occurred, and with county and year indices corresponding respectively to $c$ and $t$. This monthly weighting is a core characteristic in our model as tropical cyclone season in the U.S. runs from June to November.

Whether the indicator is measured at state or county level, we select the maximum wind speed value felt in each pixel each period, irrespective of the number of cyclone events $S$ identified within each pixel for the given year as in Felbermayr \& Gröschl (2014). This methodological choice is challenged in robustness estimations with an analysis of the impact with respect to the frequency of events.

In each case, the indicator weights wind speed levels over all pixels by the share of exposed population. Such weighting process follows recommendations to normalize natural disaster predictors in Nordhaus (2006). Ultimately, the aim is to determine the average effect of storm events on an average pixel for a given area. This measurement can also be interpreted as an estimate of a cyclone's average intensity in a randomly selected unit of land for each county, or the value of a given cyclone's intensity if this one had struck the county homogeneously across all its locations (Hsiang \& Jina, 2014). 

\subsection{Empirical strategy}
The empirical framework follows essentially Strobl (2011) and resort to analyses with spatial panel models including a lagged value for the log of personal income per capita, so that spatial correlations between states - or likewise, counties - and initial growth conditions are properly considered. More specifically, among the wide range of spatial dependence models, our specification tests for states, which are presented in Appendix~\ref{appendixA}, call for the use of spatial error model (SEM), which can be expressed as follows:
    
\begin{equation}
\label{eq3}
    g_{i,q} =  \alpha ln(Pers.Inc.pc)_{i,q-1} + \beta \overline{Cyc}_{i,q} + \mu_i + \delta_q + u_{i,q}
\end{equation}
\begin{equation*}
    u_{i,q} = \lambda \sum \limits_{i \neq j} w_{i,j} u_{j,q} + \varepsilon_{i,q}
\end{equation*}
with, $g_{i,q} = ln(Pers.Inc.pc)_{i,q} - ln(Pers.Inc.pc)_{i,q-1}$ for a state $i$ and quarter $q$. $w_{i,j}$ corresponds to an element of the spatial weighting matrix $W_N \in \mathbb{R}^{N\times N}$, which defines neighboring relationships between each of the $N$ states. $\mu_i$ and $\delta_q$ represent respectively unobserved state-specific and time-specific effects. $\lambda$ denotes the scalar spatial autoregressive coefficient, and assumed to verify $\lvert  \lambda \rvert < 1$. Finally, we assume that $\varepsilon_{i,q}$ are all independent and identically distributed with mean $0$ and finite variance $\sigma^2_{\varepsilon}$.  

As for county level estimations, specification tests leading to the choice of a SEM are rejected. Instead, results rather show that the appropriate model corresponds to a spatial autoregressive model with autoregressive disturbances (SARAR), also called spatial autocorrelated model (SAC) which is a more general class of spatial models than SEM.\footnote{See Baltagi (2021) for a detailed presentation of spatial panel models.} Thus, the equation of interest for counties corresponds to:

\begin{equation}
\label{eq4}
    g_{c,t} =  \rho \sum \limits_{c \neq j}w_{c,j} g_{j,t} + \alpha ln(Pers.Inc.pc)_{c,t-1} + \beta \overline{Cyc}_{c,t} + \mu_c + \delta_t + u_{c,t}
\end{equation}
\begin{equation*}
    u_{c,t} = \lambda \sum \limits_{c \neq j} w_{c,j} u_{j,t} + \varepsilon_{c,t}
\end{equation*}
where indices $c$ and $t$ are related to a given county and year, respectively. The dependent variable, all regressors and disturbances are defined as in the previous equation, with the exception that, now, the equation of interest includes a spatial lag $Wg= \sum \limits_{c \neq j}w_{c,j} g_{j,t}$, and as $\lambda$ its regression coefficient is also assumed to be bounded in absolute value: $\lvert \rho \rvert < 1$. 

As already stated in subsection~\ref{sec3.1}, the availability of quarterly data at state level allows for greater accuracy in the impact evaluation. More importantly, the latter point also offers the possibility to withdraw the monthly weighing that is included to $\overline{Cyc}_{c,t}$ in county-year analyses.

\begin{figure}[H]
    \centering
    \includegraphics[width = 1\textwidth]{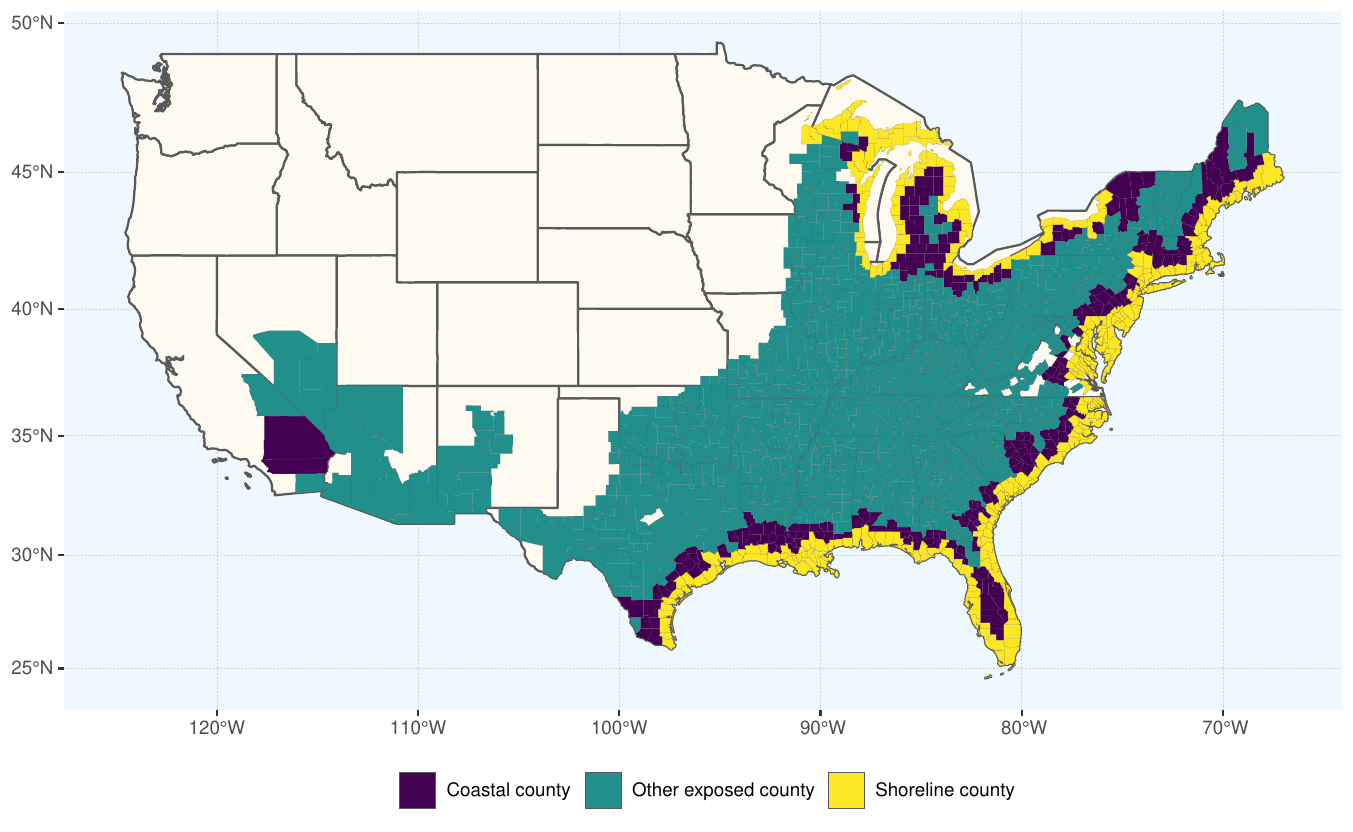}
    \caption{Categories of counties that were exposed to tropical cyclones from 1970 to 2020, given their geographical position in the U.S. and based on NOAA classification.}
    \label{fig5}
\end{figure}

In what follows, we split up our sample into three categories for states, and five categories for counties. First, we consider all 48 contiguous U.S. states (respectively all counties). Then we run regressions on the subsample of exposed states (respectively exposed counties). Third, we focus on 19 coastal states as in Strobl (2011) and all their counties that are exposed to tropical cyclones. In addition, we estimate the impact of tropical cyclones on coastal counties and on shoreline counties as defined by the NOAA. NOAA's list of coastal counties is based on NOAA coastal watersheds and U.S. Geological Survey's (USGS) coastal cataloging units. It corresponds to counties within which water flows into the ocean or Great Lakes. A coastal watershed county has at least 15 percent of its area in a coastal watershed or has a land area that accounts for at least 15 percent of a coastal cataloging unit as defined by the USGS. Shoreline counties are a subgroup of coastal counties. These counties are defined as those that are directly adjacent to the open ocean, major estuaries, and the Great Lakes.\footnote{For further information on these definitions, see \url{https://coast.noaa.gov/digitalcoast/training/defining-coastal-counties.html}.} Both of these categories are highlighted in Figure~\ref{fig5}. Table~\ref{tabSD} summarizes economic growth and cyclone intensity measurement data for each different sample. At state level, the highest cyclone measurement value is recorded in New Jersey in 2012 (Q4) when the State was affected by Hurricane Sandy (178.8 km/h). As for counties, according to our indicator, the most powerful cyclonic year is recorded in Baker County in Georgia in 2018 (196.2 km/h), followed by Cameron County in Louisiana in 2005 (186.4 km/h).

\begin{table}[H] \centering

\caption{Summary Statistics}

\begin{adjustbox}{width = 1 \textwidth}
\begin{tabular}{@{\extracolsep{5pt}} lcccccc} 
\\[-1.8ex]\hline 
\hline \\[-1.8ex] 
& & & \multicolumn{4}{c}{Average values from 1970 to 2020}  \\[+0.5ex]
            \cline{4-7} 
& Sample & Obs. & Mean & Std. Dev. & Min. & Max. \\ 
\hline \\[-1.8ex] 
\textit{State level variables} & & & & & \\[+0.5ex]
Personal income per capita growth rate ($g_{i,q}$) & All states & 9792 & 0.0038 & 0.017 & -0.25 & 0.19 \\
& Exposed states & 7548 & 0.0037 & 0.014 & -0.11 & 0.15 \\
& Coastal states & 3876 & 0.0039 & 0.013 & -0.081 & 0.11 \\~\\

Cyclone intensity measurement ($\overline{Cyc}_{i,q}$) & All states & 9792 & 2.59 & 13.7 & 0 &  171.8 \\
& Exposed states & 7548 & 3.35 & 15.5 & 0 & 171.8 \\
& Coastal states & 3876 & 5.51 & 19.9 & 0 & 171.8 \\

\hline \\[-1.8ex]
\textit{County level variables} & & & & \\[+0.5ex]
Personal income per capita growth rate ($g_{c,t}$) & All counties & 154938 & 0.016 & 0.066 & -1.36 & 1.47 \\
& Exposed counties & 104652 & 0.016 & 0.045 & -1.24 & 1.33 \\
& Coastal states' counties & 55335 & 0.017 & 0.048 & -1.24 & 1.33 \\
& NOAA coastal counties & 21573 & 0.017 & 0.042 & -1.24 & 1.06 \\
& NOAA shoreline counties & 11832 & 0.017 & 0.044 & -1.24 & 1.06 \\~\\
Cyclone intensity measurement ($\overline{Cyc}_{c,t}$) & All counties & 154938 & 3.02 & 12.4 & 0 & 196.2 \\
& Exposed counties & 104652 & 4.47 & 14.9 & 0 & 196.2 \\
& Coastal states' counties & 55335 & 6.58 & 18.0 & 0 & 196.2 \\
& NOAA coastal counties & 21573 & 9.91 & 21.7 & 0 & 186.4 \\
& NOAA shoreline counties & 11832 & 11.4 & 23.7 & 0 & 186.4 \\

\hline \\[-1.8ex] 

\end{tabular} 
\end{adjustbox}
\label{tabSD}
\end{table}

\subsection{Results at state level}
Table~\ref{tab3} outlines the results at state level. Across all our estimations on different samples, results remain statistically insignificant. Notwithstanding the absence of impact of tropical cyclones on states' economic growth, these results confirm the presence of spatial autocorrelation within our data set as the spatial error coefficient $\lambda$ is always significant at the 1\% level. 
Additional estimations that include lagged values of cyclone predictor are presented in Appendix~\ref{appendixB}. Indeed, one can rightly think that the impact of tropical cyclones might be visible only some quarters later. As such, we include up to three lags so that the impact is now estimated over a year. This change in specification does not alter the above conclusions, and the impact is consistently insignificant in the three quarters that follow the one when the cyclone makes landfall.
\begin{table}[H]
\caption{Results for regressing quarterly state growth on cyclone intensity}
    \centering
    \begin{adjustbox}{width = 0.7\textwidth}
        \begin{tabular}{lccc} 
        \hline \hline
        
             & All states & Exposed states & Coastal states \\
             & (1) & (2) & (3) \\
            \hline \\
            $ln(Pers.Inc.pc)_{i,q-1}$ & -0.034*** & -0.021*** & -0.020***  \\
             & (0.003) & (0.002) & (0.003) \\
            $\overline{Cyc}_{i,q}$  & -8.54e-07 & 6.61e-06 & 0.00001 \\
              & (0.00001) & (7.63e-06) & (8.22e-06) \\
            $\lambda$ & 0.40*** & 0.25*** & 0.17*** \\
              & (0.01) & (0.02) & (0.02) \\~\\
             
            Overall $R^2$ & 0.00 & 0.00 & 0.00 \\
            Observations & 9792 & 7548 & 3876 \\
            Number of states & 48 & 37 & 19 \\
            
        \hline \hline 
        \end{tabular}
    \end{adjustbox}

\footnotesize{\underline{Notes:} Spatial error model estimator with two-way fixed effects. Robust standard errors are in parentheses. Time and state fixed effects are included, but not reported in the table. \\
Significance levels : *** 1\% ; ** 5\% ; * 10\%. }
\label{tab3}
\end{table}

\subsection{Results at county level}

Table~\ref{tab4} presents the results from estimating equation~(\ref{eq4}), \textit{i.e.} at county level. Once again, the null hypothesis that tropical cyclones do not affect growth is accepted across all our estimations. Columns 1 to 4 show a negative but statistically insignificant relationship between cyclone intensity and economic activity on average across all U.S. counties, exposed counties, coastal states' counties, or coastal counties as defined by the NOAA. In contrast, point estimate is positive and non-significant on average across our last sample of shoreline counties. As in our estimations at state level, results confirm the presence of spatial autocorrelation as coefficients on the spatially lagged dependent variable and spatial error term are both significant at the 1\% level in all estimations, except for the sample of shoreline counties, where point estimate for $\rho$ is significant at the 5\% level. Additionally, to interpret the sign of the coefficient associated with $\rho$, as this one is negative, this means that neighboring counties have a detrimental effect on a given county's growth.

\begin{table}[H]
\caption{Results for regressing annual county growth on cyclone intensity}
    \centering
    \begin{adjustbox}{width = 1\textwidth}
        \begin{tabular}{lccccc} 
        \hline \hline
        
             & All counties & Exposed counties & Coastal states' & Coastal counties & Shoreline counties \\
             & & & counties & (NOAA definition) & (NOAA definition)\\
             & (1) & (2) & (3) & (4) & (5) \\
            \hline \\
            $ln(Pers.Inc.pc)_{c,t-1}$ & -0.182***  & -0.130***  & -0.137*** & -0.108*** & -0.122*** \\
             & (0.001) & (0.002) & (0.002)  & (0.003) & (0.004) \\
            $\overline{Cyc}_{c,t}$  & -6.38e-06 & -0.00001 & -0.00001 & -0.00002 & 5.48e-07 \\
              & (0.00001) & (0.00001) & (0.00002) & (0.00002) & (0.00002) \\
            $\rho$ & -0.093*** & -0.073*** & -0.066*** & -0.084*** & -0.025** \\
              & (0.001) & (0.002) & (0.002) & (0.004) & (0.011) \\
            $\lambda$ & 0.138*** & 0.127*** & 0.121*** & 0.131*** & 0.103*** \\
              & (0.0003) & (0.0007) & (0.0008) & (0.002) & (0.008) \\~\\
             
            Overall $R^2$ & 0.00 & 0.00 & 0.00 & 0.00 & 0.01 \\
            Observations & 154938 & 104652 & 55335 & 21573 & 11832 \\
            Number of counties & 3038 & 2052 & 1085 & 423 & 232 \\
            
        \hline \hline 
        \end{tabular}
    \end{adjustbox}

\footnotesize{\underline{Notes:} Spatial autocorrelated model estimator with two-way fixed effects. Robust standard errors are in parentheses. Time and county fixed effects are included, but not reported in the table. \\
Significance levels : *** 1\% ; ** 5\% ; * 10\%. }
\label{tab4}
\end{table}

\subsection{Robustness checks}
This section considers a variety of robustness checks. Analyses are of two kinds. First, we keep the empirical strategy used above and consider robustness to alternative cyclone predictors. More specifically, on the one hand, we investigate an alternative formulation of our cyclone indicator that only keeps the share of exposed population as in Krichene \& al. (2021). On the other hand, we check whether baseline results are altered when examining the effects of the frequency of cyclone events instead of our wind speed intensity measurement. Second, we reconsider the main specification and report the results when using a fixed effects estimator with Conley (1999) standard errors. These standard errors are robust to cross-sectional spatial dependence and serial correlation below some fixed cutoff values which are here set to 1000km and 3 time lags respectively. Hence, this alternative estimator provides a strong complement to our baseline spatial panel models estimations in which neighboring relationship is defined by sharing common borders. Results for each different state or county sample are summarized in Table~\ref{tab5} and Table~\ref{tab6}.

\subsubsection{State level}
Overall, we find that using alternative indicators or changing the estimator does not change our conclusions. Removing the wind speed parameter from our baseline indicator does not change the sign of the relationship but increases the magnitude of coefficients. The quarterly number of cyclone events has a positive contemporaneous impact on average across all different samples, \textit{i.e.} across all states, exposed states and coastal states. Finally, estimating equation~(\ref{eq3}) by applying a fixed effects estimator with Conley (1999) standard errors consistently displays negative immediate effects. However, in no case are these results statistically significant, meaning that state growth is arguably unaffected by tropical cyclone events in the quarter they occur.   

\begin{table}[H]
\caption{Robustness checks results: regressions on quarterly state growth}
    \centering
    \begin{adjustbox}{width = 1\textwidth}
        \begin{tabular}{lcccccc} 
        \hline \hline
        
             & All states & Exposed states & Coastal states \\
             & (1) & (2) & (3) \\
            \hline 
            \small \textit{Alternative cyclone indicator: share of exposed population} & & & \\~\\
            $ln(Pers.Inc.pc)_{i,q-1}$ & -0.034*** & -0.021*** & -0.020*** \\
            & (0.003) & (0.002) & (0.002) \\
            $Pop.~exposed_{i,q}$  & -0.00004 & 0.0005 & 0.001 \\
            & (0.00100) & (0.0007) & (0.0008) \\
            $\lambda$ & 0.40*** & 0.25*** & 0.17*** \\
            & (0.01) & (0.02) & (0.02) \\~\\
            
            Overall $R^2$ & 0.00 & 0.00 & 0.00 \\
            Observations & 9792 & 7548 & 3876 \\
            Number of states & 48 & 37 & 19 \\
            \hline
            \small \textit{Alternative cyclone indicator: quarterly number of cyclone events} & & & \\~\\
            $ln(Pers.Inc.pc)_{i,q-1}$ & -0.034*** & -0.021*** & -0.020*** \\
            & (0.003) & (0.002) & (0.002) \\
            $Nb.~Cyc_{i,q}$  & 0.0002 & 0.0004 & 0.0003 \\
            & (0.0004) & (0.0003) & (0.0003) \\
            $\lambda$ & 0.40*** & 0.25*** & 0.17*** \\
            & (0.01) & (0.02) & (0.02) \\~\\
            
            Overall $R^2$ & 0.00 & 0.00 & 0.00 \\
            Observations & 9792 & 7548 & 3876 \\
            Number of states & 48 & 37 & 19 \\
            \hline
            \small \textit{Fixed effects estimator using Conley 1999 standard errors} & & & & \\~\\
            $ln(Pers.Inc.pc)_{i,q-1}$ & 0.0004*** & 0.0004*** & 0.0004*** \\
            & (0.00004) & (0.00004) & (0.00005) \\
            $\overline{Cyc}_{i,q}$  & -0.00002 & -0.00002 & -0.00002 \\
            & (0.00003) & (0.00003) & (0.00003) \\~\\
            $R^2$ & 0.05 & 0.07 & 0.08 \\
            Observations & 9792 & 7548 & 3876 \\
            Number of states & 48 & 37 & 19 \\
            
        \hline \hline 
        \end{tabular}
    \end{adjustbox}

\footnotesize{\underline{Notes:} The two first sections of the table report regressions that use spatial error model estimator with two-way fixed effects. Robust standard errors are in parentheses. Time and state fixed effects are included, but not reported in the table. Bottom section presents OLS regression results with standard errors corrected for cross-sectional spatial dependence and panel-specific serial correlation as in Conley (1999). The spatial correlation cutoff is 1000km, and the serial correlation one is 3 lags. \\
Significance levels : *** 1\% ; ** 5\% ; * 10\%. }
\label{tab5}
\end{table}

\subsubsection{County level}
Table~\ref{tab6} reports the results for counties. Point estimates for the share of exposed population are positive across all estimations. The annual number of cyclone events has always a negative effect, except for the estimation on NOAA coastal counties. Estimations with Conley (1999) standard errors show a positive effect in the year the cyclone occurs across all samples. However, as none of the coefficients associated with our cyclone indicators are significant, the absence of impact is confirmed once again by these estimations. 

\begin{table}[H]
\caption{Robustness checks results: regressions on annual county growth}
    \centering
    \begin{adjustbox}{width = 1\textwidth}
        \begin{tabular}{lccccc} 
        \hline \hline        
             & All counties & Exposed counties & Coastal states' & Coastal counties & Shoreline counties \\
             & & & counties & (NOAA definition) & (NOAA definition)\\
             & (1) & (2) & (3) & (4) & (5) \\
            \hline
            \small \textit{Alternative cyclone indicator: share of exposed population} & & & & \\~\\
            $ln(Pers.Inc.pc)_{c,t-1}$ & -0.182*** & -0.130*** & -0.136*** & -0.108*** & -0.122***\\
             & (0.001) & (0.002) & (0.002) & (0.003) & (0.004) \\
            $Pop.~exposed_{c,t}$  & 0.0003 & 0.0001 & 0.0002 & 0.0002 & 0.0003 \\
              & (0.0003) & (0.0002) & (0.0003) & (0.0004) & (0.0005) \\
            $\rho$ & -0.093*** & -0.073*** & -0.066*** & -0.083*** & -0.025** \\
              & (0.001) & (0.002) & (0.002) & (0.004) & (0.011) \\
            $\lambda$ & 0.138*** & 0.127*** & 0.121*** & 0.131*** & 0.103*** \\
              & (0.0003) & (0.0007) & (0.001) & (0.002) & (0.008) \\~\\
            Overall $R^2$ & 0.00 & 0.00 & 0.00 & 0.00 & 0.01 \\
            Observations & 154938 & 104652 & 55335 & 21573 & 11832 \\
            Number of counties & 3038 & 2052 & 1085 & 423 & 232 \\
            \hline
            \small \textit{Alternative cyclone indicator: yearly number of cyclone events} & & & & \\~\\
            
            $ln(Pers.Inc.pc)_{c,t-1}$ & -0.182*** & -0.130*** & -0.136*** & -0.108*** & -0.122***\\
            & (0.001) & (0.002) & (0.002) & (0.003) & (0.004) \\
            $Nb.~Cyc_{c,t}$  & -0.0008 & -0.0007 & -0.0003 & 0.0003 &  -0.0002 \\
              & (0.0008) & (0.0006) & (0.0006) & (0.0008) & (0.0009) \\
            $\rho$ & -0.093*** & -0.072*** & -0.066*** & -0.083*** & -0.025** \\
              & (0.001) & (0.0006) & (0.002) & (0.004) & (0.011) \\
            $\lambda$ & 0.138*** & 0.127*** & 0.121*** & 0.131*** & 0.103*** \\
              & (0.0003) & (0.0007) & (0.001) & (0.002) & (0.008) \\~\\
            Overall $R^2$ & 0.02 & 0.00 & 0.00 & 0.00 & 0.01 \\
            Observations & 154938 & 104652 & 55335 & 21573 & 11832 \\
            Number of counties & 3038 & 2052 & 1085 & 423 & 232 \\
            \hline
            \small \textit{Fixed effects estimator using Conley 1999 standard errors} & & & & \\~\\
            $ln(Pers.Inc.pc)_{c,t-1}$ & 0.002*** & 0.002*** & 0.002*** & 0.001*** & 0.001*** \\
             & (0.0002) & (0.0002) & (0.0002) & (0.0002) & (0.0002) \\
            $\overline{Cyc}_{c,t}$  & 0.00003 & 0.00003 & 0.00003 & 0.00005 &  0.00005 \\
              & (0.00005) & (0.00004) & (0.00004) & (0.00004) & (0.00004) \\~\\
            $R^2$ & 0.05 & 0.11 & 0.11 & 0.14 & 0.13 \\
            Observations & 154938 & 104652 & 55335 & 28203 & 15246 \\
            Number of counties & 3038 & 2052 & 1085 & 423 & 232 \\
            
        \hline \hline 
        \end{tabular}
    \end{adjustbox}

\footnotesize{\underline{Notes:} The two first sections of the table report regressions that use spatial autocorrelated model estimator with two-way fixed effects. Robust standard errors are in parentheses. Time and county fixed effects are included, but not reported in the table. Bottom section presents OLS regression results with standard errors corrected for cross-sectional spatial dependence and panel-specific serial correlation as in Conley (1999). The spatial correlation cutoff is 1000km, and the serial correlation one is 3 lags. \\
Significance levels : *** 1\% ; ** 5\% ; * 10\%.}
\label{tab6}
\end{table}

\section{Focusing on Florida}
\label{sec4}
So far, our results show that tropical cyclones do not impact economic activity in the quarter they occur in U.S. states, but also in the year they occur in U.S. counties. Having said that, in a second phase, we wonder if our main results imply that there is no effect either if we address our research question at a smaller scale.

In particular, we now focus on Florida, which is the most frequently exposed state in the U.S. from 1970 to 2020 with 83 cyclones (Figure~\ref{fig1}). This feature is all the more important, given that the following most frequently hit states are North Carolina with 65 events recorded during the sample period, then Louisiana with 54 tropical cyclones and Texas with 53. Hence, there is quite a large gap in the exposure to these catastrophes between Florida and its U.S. counterparts. Several other reasons might be put forward to justify a focus on this specific state when studying cyclonic risk. Florida was the fourth richest state in the U.S. at the end of 2020, behind the States of New York, Texas and California according to our personal income data. It is also the third most populated State of the U.S. in 2020 behind Texas and California. In addition, one striking feature in Florida is the concentration of most of its largest metropolitan areas along the shoreline, or at least, considered as coastal according to the NOAA. Indeed, Miami, which is the 8th largest metropolitan area in the U.S. in 2020 with 6.1 million inhabitants, Tampa (18th largest), Orlando (22nd largest) and Jacksonville (39th largest) are all located near U.S. coasts, and arguably more likely to face higher cyclone intensities compared to other areas that would be located deeper in lands.\footnote{Population size rankings come from U.S. Census Bureau data. For further information, see \url{https://www.census.gov/data/tables/time-series/demo/popest/2020s-total-metro-and-micro-statistical-areas.html}.} Notably, no other state exposed to tropical cyclones shows such geographic or demographic characteristics in our sample. Third, Florida's increased vulnerability to tropical cyclones due to this high concentration of population along the coasts is emphasised by a strong dependence on tourism activity. In fact, tourism represents the first source of revenue of the State. Agricultural activity, which is also at risk in the context of tropical cyclones, represented 1,2\% of Floridian GDP in 2020 (United States Department of Agriculture, 2021). Hence, agriculture also brings a non-negligible contribution to Florida State's economy. Finally, as Florida is at the edge of the Caribbean, its local climate is more prone to higher cyclone exposure and intensities as it is surrounded by warmer water temperatures compared to other states due to its greater proximity to the equator. Florida also contains main harbours that export towards Latin America or the Caribbean. For all these reasons, several studies on the economic impact of tropical cyclones have chosen to focus on Florida (Belasen \& Polachek, 2009; Brown \& al., 2021; Pollack \& Kauffmann, 2022).

\subsection{State level analysis}

To estimate the impact of tropical cyclones on Floridian economic growth, we opt for an Auto-Regressive Integrated Moving Average with exogenous factors (ARIMAX) model and follow the  Box \& Jenkins (1976) time series modeling approach. This method consists in a three-stage procedure: model identification, parameter estimation, and diagnostic checking. In the present case, the equation of interest is expressed as:

\begin{equation}
    g_t = A(L)g_{t-1} + B(L)\overline{Cyc}_{t} + C(L)\varepsilon_t
    \label{eqflorida1}
\end{equation}

\noindent with $A(L), B(L)$ and $C(L)$ polynomials in the lag operator $L$, $g$ (deseasonalized) personal income per capita growth, $\overline{Cyc}_{t}$ the usual population-weighted average of the maximum sustained wind speed over all pixels in Florida each quarter~$t$, and $\varepsilon$ the white noise error term, independent and identically distributed.

\begin{figure}[H]
    \centering
    \includegraphics[width = 0.49 \textwidth]{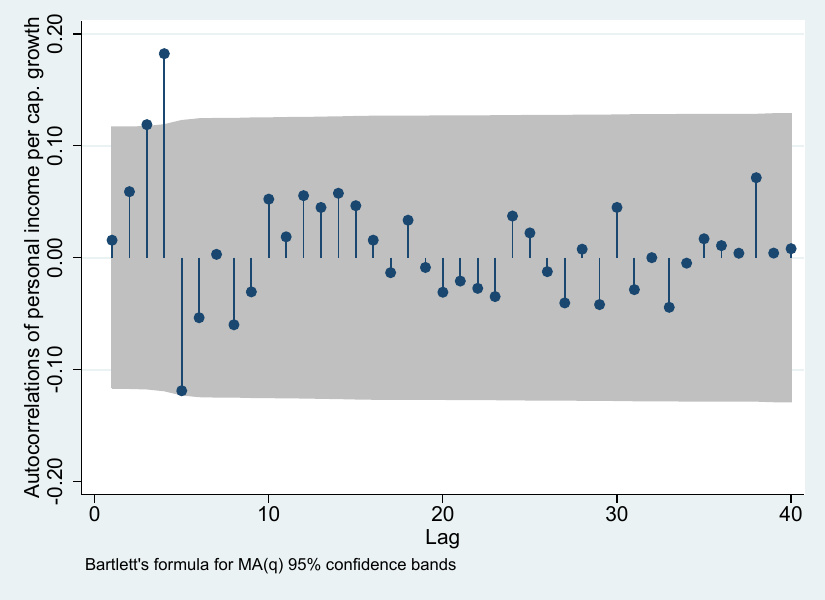}
    \includegraphics[width = 0.49 \textwidth]{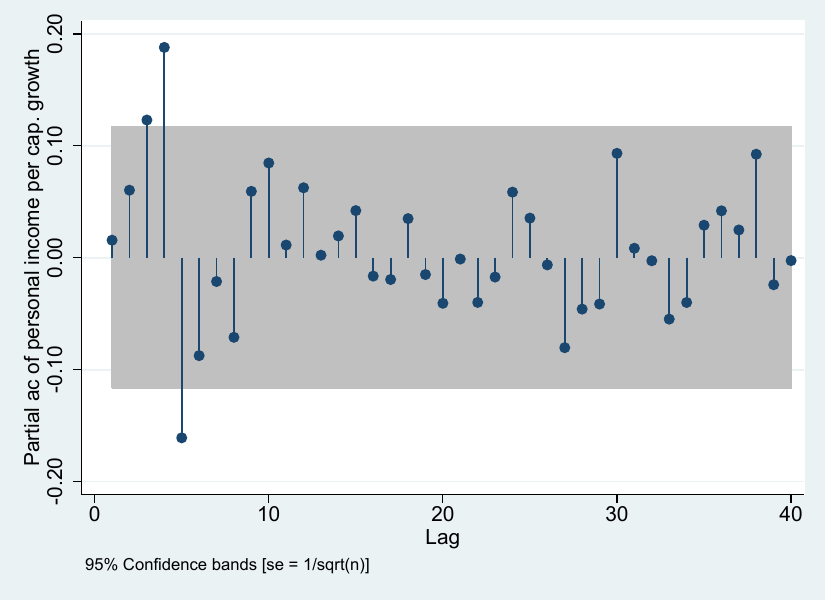}
    \caption{Autocorrelation function (left) and partial autocorrelation function (right) of Floridian deseasonalized personal income par capita growth.}
    \label{fig6}
\end{figure}

The identification step relies on an examination of the time path and the autocorrelation function (ACF) of Floridian series in order to get information about stationarity, structural breaks, trends, or seasonality. This step corresponds to the analyses presented at the end of section~\ref{sec2}, and led to the choice of $g_t$. Floridian growth's ACF and PACF suggest that the model can be either autoregressive (AR) as point estimates are alternating positive and negative, or a mixed autoregressive and moving average (ARMA) model as the decay is starting after a couple of lags, and more specifically, here, five lags (Figure~\ref{fig6}).

Then, the second step is to estimate the model. To do this, the optimal number of lags to be included in the model can be determined using the ACF plot of $g_t$. The latter plot recommends a number of lags comprised between 3 and 5. Another alternative is to perform optimal lag tests. In this paper, four of them are selected, namely, the Final Prediction Error (FPE), Akaike information criterion (AIC), Hannan-Quinn criterion (HQIC) and Schwarz information criterion (SIC). The estimated optimal lag length corresponds to the one minimizing the criterion. The advantage of these optimal lag tests is to include the exogenous variable in the estimation. Table~\ref{tab7} presents the results of the tests. FPE, HQIC and SIC all suggest 5 lags, which somehow joins the interpretations made with the ACF plot, while SIC suggests that no lag should be included. Hence, considering altogether the results of the ACF, the PACF and those of the optimal lag tests, we choose to estimate the impact with 5 lags.

\begin{table}[H]
    \centering
    \caption{Optimal lag tests for personal income per capita growth}
    \begin{adjustbox}{width=1\textwidth}
        \begin{tabular}{lcccc} 
        \hline \hline
             & Final Prediction Error & Akaike information criterion  & Hannan-Quinn criterion & Schwarz information criterion \\
              & (FPE) & (AIC) & (HQIC) & (SIC) \\
            \hline
            Nb lags & & & & \\
    0  &  .000148 & -5.97863 & -5.96739  & \textbf{-5.9507}* \\ 
    1    &  .000148 & -5.97778 & -5.96092 & -5.93588 \\ 
    2   &  .000148 & -5.98358  & -5.9611 & -5.92771  \\
    3   &  .000147 & -5.98817 & -5.96008 & -5.91834  \\
    4    &  .000145  & -6.0039 & -5.97019 & -5.92011   \\
    5   &  \textbf{.000143}* & \textbf{-6.01646}* & \textbf{-5.97712}* & -5.91869 \\
    6   &  .000144 & -6.00882 & -5.96387  & -5.8971   \\
    7   &   .000145 & -6.00165 & -5.95108 & -5.87596   \\
    8   &  .000145  & -6.00022 & -5.94403 & -5.86056  \\
    9   &  .000146 & -5.99685 & -5.93504 & -5.84322  \\

        \hline \hline 
        \end{tabular}
    \end{adjustbox}
\footnotesize \underline{Notes:} * denotes the optimal number of lags to be included according to a given criterion. Estimations with a maximum number of lags fixed at 30. $\overline{Cyc}_{i,q}$ exogenous variable.
\label{tab7}
\end{table}

Finally, Table~\ref{tab8} outlines the results. Across all estimations, Floridian economic activity is found to be adversely affected one quarter after a cyclone strike, and this negative impact is muted afterwards with most of the specifications. According to the AIC and BIC criteria, the ARIMAX $(5,0,2)$ model seems to provide the best fit. A one km/h increase in the cyclone intensity measurement is responsible for 0.005 percentage points of growth loss in the quarter the tropical cyclone strikes Florida (significant at the 10\% level). The point estimate two quarters later is also found to be negative and significant at the 10\% level (-0.006 percentage points). For all the regressions, the invertibility condition is verified, meaning that all the roots of polynomials in lag operators lie inside the unit circle. 

In order to prove that there is a differential impact across states, and that high exposure to tropical cyclones is not the only criterion that is responsible for a local negative impact, in Appendix~\ref{appendixC}, we examine the results using the same methodology for North Carolina, Louisiana and Texas. The preferred specification according to AIC and BIC criteria is ARIMAX(4,0,0) for North Carolina, ARIMAX(1,0,2) for Louisiana, and ARIMAX(1,0,0) for Texas. In no case a significant impact is found for these three states. 

\begin{table}[H]
\caption{Results for regressing quarterly growth on cyclone indicator}
    \centering
    \begin{adjustbox}{width= 0.7\textwidth}
        \begin{tabular}{lccccc} 
        \hline \hline
              & ARIMAX & ARIMAX  & ARIMAX & ARIMAX & ARIMAX  \\
              & $(5,0,0)$ & $(5,0,1)$ & $(5,0,2)$ & $(5,0,3)$ & $(5,0,4)$  \\
            \hline
    $\overline{Cyc}_q$  & -0.00004 & -0.00006*  & \textbf{-0.00005*} & -0.00005*  & -0.00005* \\
      &  (0.00003) & (0.00003) & (0.00003) &  (0.00003)  &  (0.00003) \\
    $\overline{Cyc}_{q-1}$ & 0.00002 & -5.02e-06  & -7.58e-07 & 3.29e-06 & -1.45e-06  \\
                &  (0.00003) & (0.00003) & (0.00003) &  (0.00004) &  (0.00004) \\
    
    $\overline{Cyc}_{q-2}$ &  -0.00004 &-0.00007* & \textbf{-0.00006*} & -0.00006 & -0.00006 \\
                & (0.00004)  & (0.00004) & (0.00004) &  (0.00004) &  (0.00004)\\
    
    $\overline{Cyc}_{q-3}$ & -0.00002 & -0.00003 & -0.00003 & -0.00005 & -0.00004 \\
                & (0.00004) & (0.00003) & (0.00004) &  (0.00004) &  (0.00004) \\
    $\overline{Cyc}_{q-4}$  & -0.00001 & -0.00003 & -0.00003  & -0.00003 & -0.00003 \\
      & (0.00003) & (0.00006) & (0.00003) &  (0.00003) &  (0.00003) \\
      
    $\overline{Cyc}_{q-5}$   & 7.16e-06 & -9.35e-06 & -7.03e-06 & -1.02e-06 & 2.10e-06 \\
      & (0.00003) & (0.00004) & (0.00003) &  (0.00003) &  (0.00003) \\
    
    \hline
    AIC  & -1201.304 & -1203.568 & \textbf{-1205.651} &  -1204.697 & -1201.396 \\
    BIC  & -1158.169 & -1157.114 & \textbf{-1159.198} & -1154.925 & -1144.988  \\

        \hline \hline 
        \end{tabular}
    \end{adjustbox}
\label{tabflorida}

\footnotesize \underline{Notes:} robust standard errors in parentheses. \\
* significant at 10 \%; ** significant at 5 \%; *** significant at 1 \%
\label{tab8}
\end{table}

\subsection{County level analysis}

Results at county level are shown in Table~\ref{tab9}. We decide to include 3 lags of the cyclone indicator in order to provide insights about longer-term effects. When using our cyclone intensity measurement $\overline{Cyc}_{c,t}$, the impact is consistently negative and significant over three years. The estimated immediate effect is -0.005 percentage points of growth for each km/h increase in storm intensity, -0.002 one year later, and then -0.003 two years later. The impact after three years is insignificant. If tropical cyclones are proxied by the share of the exposed population over each county, the immediate impact is comparable in magnitude but loses its statistical significance. However, the impact is larger in magnitude with better statistical significance one and two years after the occurrence of the catastrophe. Finally, the frequency of events in Floridian counties has no impact, whether in the short- or long run according to our model. 
As in the last subsection, we replicate the estimations for North Carolinian, Louisianian, and Texan exposed counties with our main cyclone intensity measurement. Point estimates are still mostly insignificant, except one and two years later in North Carolina, for which a positive impact is found (+0.002 and +0.001 percentage points, respectively) and the second cyclone lag's point estimate for Texas (+0.004 percentage points). These positive and significant delayed impacts suggest the efficiency of recovery processes after a cyclone strike, even though the immediate impact is, in each case, insignificant. This might also mean that tropical cyclones actually stimulate local economic activity one and two years later. 

\begin{table}[H]
\caption{Results for regressing annual Floridian counties' growth on cyclone variables}
    \centering   
    \begin{adjustbox}{width = 0.8\textwidth}
        \begin{tabular}{lccc}
        \hline \hline
          & Independent variable & Independent variable & Independent variable \\
          & $x = \overline{Cyc}_{c,t}$ & $x = Pop.~Exposed_{c,t}$ & $x = Nb.~Cyc_{c,t}$ \\
          & (1) & (2) & (3) \\
            \hline
             $ln(Pers.Inc.pc)_{i,t-1}$ & -0.047*** & -0.047*** & -0.046*** \\
             & (0.005)  & (0.005) &  (0.005) \\
            $x_{c,t}$  & -0.00005* & -0.00005 & -0.0007 \\
              & (0.00004) & (0.00007) & (0.0007) \\
            $x_{c,t-1}$  & -0.00002* & -0.002*** &  0.0002 \\
              & (0.00001) & (0.0008) & (0.0008) \\
            $x_{c,t-2}$  & -0.00003*** & -0.003*** & -0.002 \\
              & (9.85e-06) & (0.0008) & (0.0008) \\
            $x_{c,t-3}$  &  6.09e-06 & -0.0003 & -0.0009 \\
              & (9.27e-06) & (0.0008) & (0.0008) \\
            $\rho$ & 0.106*** & 0.105*** & 0.106*** \\
              & (0.007) & (0.007) & (0.007) \\
            $\lambda$ & -0.088*** & -0.086*** & -0.089*** \\
              & (0.012) & (0.013) & (0.011) \\~\\
              
            Overall $R^2$ & 0.00 & 0.00 & 0.00 \\
            Observations & 3283 & 3283 & 3283 \\
            Number of counties & 67 & 67 & 67 \\
                        
        \hline \hline 
        \end{tabular}
    \end{adjustbox}

\footnotesize{\underline{Notes:} Spatial error model estimator with two-way fixed effects. Robust standard errors are in parentheses. Time and county fixed effects are included, but not reported in the table. \\
Significance levels : *** 1\% ; ** 5\% ; * 10\%. }
\label{tab9}
\end{table}

\section{Conclusion}
\label{sec5}

Tropical cyclones are fast onset and hardly predictable disasters, which have a tremendous potential to disrupt societies. This paper evaluates their impact on U.S. economic growth at county and state levels. Using $0.1^\circ$ latitude $\times$ $0.1^\circ$ longitude pixel level cyclone data, we build a measurement of disaster intensity which include both wind speed and share of exposed population. Data on personal income per capita come from the Bureau of Economic Analysis, which releases annual data at county level and quarterly data at state level. As the economic activity of a given area is undoubtedly influenced by those of its neighbours, our panel data estimations consider spatial correlations across groups, which are incorporated either by using spatial error models or spatial autocorrelated models. 

Our first strand of results suggests that tropical cyclones do not influence economic activity in the U.S. when counties or states are sampled all over the country. However, in its second part, this study focuses on Florida State and its counties. Florida presents several factors of vulnerability to tropical cyclones. In particular, it is the most frequently exposed state in the U.S. Analyses are conducted using time series modeling for the single-state investigation, while the econometric strategy remains unchanged for the estimations within counties. This last series of estimations unveils significant results of growth depletion. On the quarter the cyclone occurs, the effect of a one km/h increase in intensity is to decrease contemporaneous growth rate by 0.005 percentage points, and by 0.006 more two quarters later. At county level, we estimate that such an increase in cyclone intensity is responsible for 0.005 percentage points in growth reduction on the year the cyclone occurs, and this negative effect remains significant up to two years after the strike. 

Hence, our findings add to the recent body of literature that claims for regional or local investigations in environmental economics as we demonstrate the heterogeneity of causal effects within the same country in the context of tropical cyclones. Further research on economic channels such as sectoral growth is required to better understand the observed growth reduction in Florida and might bring more external validity to the present results.

\newpage 

\Large{\textbf{References}}
\\~\\
\normalsize

\noindent Baltagi, B. H. (2021). Econometric Analysis of Panel Data. Springer Texts in Business and Economics, Springer, edition 6, number 978-3-030-53953-5, March. \\~\\
Belasen, A. R., \& Polachek, S. W. (2009). How disasters affect local labor markets the effects of hurricanes in Florida. Journal of Human Resources, 44(1), 251-276.\\~\\
Box, G. E. P. \& Jenkins, G. M. (1976). Time Series Analysis: Forecasting and Control, revised ed. San Francisco: Holden-Day. \\~\\
Brown, C. E., Alvarez, S., Eluru, N., \& Huang, A. (2021). The economic impacts of tropical cyclones on a mature destination, Florida, USA. Journal of Destination Marketing \& Management, 20, 100562.\\~\\
Conley, T. G. (1999). GMM estimation with cross sectional dependence. \textit{Journal of econometrics, 92}(1), 1-45.\\~\\
Felbermayr, G., \& Gröschl, J. (2014). Naturally negative: The growth effects of natural disasters. \textit{Journal of development economics, 111}, 92-106.\\~\\
Geiger, T., Frieler, K., \& Bresch, D. N. (2018). A global historical data set of tropical cyclone exposure (TCE-DAT). Earth Syst. Sci. Data, 10, 185–194, https://doi.org/10.5194/essd-10-185-2018. \\~\\
Hsiang, S. M., \& Jina, A. S. (2014). The causal effect of environmental catastrophe on long-run economic growth: Evidence from 6,700 cyclones (No. w20352). National Bureau of Economic Research.\\~\\
IPCC, 2022: \textit{Climate Change 2022: Impacts, Adaptation, and Vulnerability.} Contribution of Working Group II to the Sixth Assessment Report of the Intergovernmental Panel on Climate Change [H.-O. Pörtner, D.C. Roberts, M. Tignor, E.S. Poloczanska, K. Mintenbeck, A. Alegría, M. Craig, S. Langsdorf, S. Löschke, V. Möller, A. Okem, B. Rama (eds.)]. Cambridge University Press. In Press.\\~\\
Krichene, H., Geiger, T., Frieler, K., Willner, S. N., Sauer, I., Otto, C. (2021). Long-term impacts of tropical cyclones and fluvial floods on economic growth – Empirical evidence on transmission channels at different levels of development. \textit{World Development}, Volume 144, 105475.\\~\\
Knapp, K. R., Kruk, M. C., Levinson, D. H., Diamond, H. J., \& Neumann, C. J. (2010). The international best track archive for climate stewardship (IBTrACS) unifying tropical cyclone data. \textit{Bulletin of the American Meteorological Society, 91}(3), 363-376.\\~\\
Nordhaus, W. D. (2006). The economics of hurricanes in the United States. \textit{National Bureau of Economic Research Working Paper Series, n°12813}\\~\\
Noy, I. (2009). The macroeconomic consequences of disasters. \textit{Journal of Development economics, 88}(2), 221-231.\\~\\
Pollack, A. B., \& Kaufmann, R. K. (2022). Increasing storm risk, structural defense, and house prices in the Florida Keys. Ecological Economics, 194, 107350.\\~\\
Strobl, E. (2011). The economic growth impact of hurricanes: Evidence from US coastal counties. \textit{Review of Economics and Statistics, 93}(2), 575-589.\\~\\

\newpage

\Large{\textbf{APPENDIX}}

\appendix
\normalsize
\setcounter{table}{0}
\renewcommand{\thetable}{A.\arabic{table}}
\section{Specification test}
\label{appendixA}
Spatial dependence models are relevant models in the presence of cross-section of countries, regions, states, or more generally, any geographic units. In fact, introducing spatial effects in panel models leads models to take into consideration the interdependence between groups. It also allows to estimate spillover effects or externalities. These models are particularly attractive when dealing with regional science or urban economics issues. One can write a general equation for spatial panel data models as:    
\begin{equation*}
    y_{i,t} =  \alpha + \rho \sum \limits_{i \neq j}w_{i,j} y _{j,t} + \beta x _{i,t} + \theta \sum \limits_{i \neq j}w_{i,j} x_{j,t} + \mu_i + \delta_t + u_{i,t}
\end{equation*}
\begin{equation*}
    u_{i,t} = \lambda \sum \limits_{i \neq j} w_{i,j} u_{j,t} + \varepsilon_{i,t}
\end{equation*}
with $y_{i,t}$ dependent variable, $x_{i,t}$ explanatory variable, and $w_{i,j}$ element of the spatial weighting matrix $W_N \in \mathbb{R}^{N\times N}$ that defines the neighboring links between each of the $N$ groups. 

Based on this general equation, we can define the following models:

 \begin{itemize}
     \item Spatial Autoregressive Model (SAR), if $\lambda = \theta = 0$.
     \item Spatial Durbin Model (SDM), if $\lambda = 0$.
     \item Spatial Autoregressive Model with Auto Regressive disturbances (SARAR), which are also called spatial autocorrelated model (SAC), if $\theta = 0$.
     \item Spatial Error Model (SEM), if $\rho = \theta = 0$.
 \end{itemize}
 
Apart from these four basic models, many other specifications can be defined such as those with $\mu_i$ random and depending on spatial lags, dynamic models, etc. 

In order to find the most appropriate model, some specification tests can be run using parametric statistical measures such as Wald tests. For instance:
\begin{itemize}
     \item After estimating a SDM, one can check whether $\theta = 0$. If the null hypothesis is accepted, then it means that in fact, the appropriate model is a SAR. 
     \item After estimating a SDM, one can check whether $\theta = -\beta \rho$. In this case, not rejecting the null hypothesis calls for a SEM model.
\end{itemize}

Table~\ref{tabA1} examines the results obtained for states with a SDM. In each case, the coefficients associated with the spatial lags of regressors are always insignificant, and the equality $\theta = -\beta \rho$ is verified. Altogether, this means that the correct model for state level analyses is the SEM. However, as for county level estimations, all $\theta = -\beta \rho$ Wald-type tests are rejected (p-value < 0.10), and spatially weighted independent variables' coefficients are found to be zero. Hence, we exploit a generalization of the SAR model, namely the SAC model. \\~\\

\begin{table}[H]
\caption{Results for regressing quarterly State growth on cyclone intensity using a Spatial Durbin Model}
\label{tabA1}
    \centering
    \begin{adjustbox}{width = 0.63\textwidth}
        \begin{tabular}{lccc} 
        \hline \hline
        
             & All states & Exposed states & Coastal states \\
             & (1) & (2) & (3) \\
            \hline \\
            $Wg$ & 0.398*** & 0.252*** & 0.146***  \\
             & (0.012) & (0.015) & (0.022) \\
            
            $ln(Pers.Inc.pc)_{i,t-1}$ & -0.038*** & -0.025*** & -0.019***  \\
             & (0.003) & (0.003) & (0.003) \\
            $\overline{Cyc}_{i,t}$  & 5.93e-07 & 2.69e-06 & 9.49e-06 \\
              & (0.00001) & (9.86e-06) & (0.00001) \\
            $W\overline{Cyc}$ & -1.00e-06 & 4.68e-06 & 6.90e-06 \\
              & (0.00002) & (0.00001) & (0.00002) \\
            
            Observations & 9792 & 7548 & 3876 \\
            Number of states & 48 & 37 & 19 \\
            Overall $R^2$ & 0.00 & 0.00 & 0.00 \\~\\

            Wald-type test $P>\chi^2$ & 0.958 & 0.263 & 0.851 \\
            $\theta$ = -$\rho \beta$ & & & \\
        \hline \hline 
        \end{tabular}
    \end{adjustbox}

\footnotesize{\underline{Notes:} Spatial Durbin Model estimator with two-way fixed effects. Robust standard errors are in parentheses. Time and state fixed effects, as well as lagged per capita income spatial lags are included, but not reported in the table. \\
Significance levels : *** 1\% ; ** 5\% ; * 10\%. }
\end{table}

\setcounter{table}{0}
\renewcommand{\thetable}{B.\arabic{table}}
\section{Additional estimations}
\label{appendixB}

\begin{table}[H]
\caption{Results for regressing quarterly state growth on cyclone intensity using a specification that includes lags}
    \centering
    \begin{adjustbox}{width = 0.6\textwidth}
        \begin{tabular}{lccc} 
        \hline \hline
        
             & All states & Exposed states & Coastal states \\
             & (1) & (2) & (3) \\
            \hline \\
            $ln(Pers.Inc.pc)_{i,q-1}$ & -0.034*** & -0.021*** & -0.020***  \\
             & (0.003) & (0.002) & (0.003) \\
            $\overline{Cyc}_{i,q}$  & 1.12e-06 & 6.61e-06 & 0.00001 \\
              & (0.00001) & (7.64e-06) & (8.21e-06) \\
            $\overline{Cyc}_{i,q-1}$  & -2.35e-06 & 3.12e-06 & 9.17e-06 \\
              & (0.00001) & (7.73e-06) & (8.31e-06) \\
            $\overline{Cyc}_{i,q-2}$  & 0.00001 & 5.32e-06 & 7.85e-06 \\
              & (0.00001) & (7.64e-06) & (8.23e-06) \\
            $\overline{Cyc}_{i,q-3}$  & 2.97e-06 & 5.82e-07 & -2.03e-06 \\
              & (0.00001) & (7.69e-06) & (8.23e-06) \\
            $\lambda$ & 0.40*** & 0.25*** & 0.17*** \\
              & (0.01) & (0.02) & (0.02) \\~\\
             
            Overall $R^2$ & 0.00 & 0.00 & 0.00 \\
            Observations & 9792 & 7548 & 3876 \\
            Number of states & 48 & 37 & 19 \\
            
        \hline \hline 
        \end{tabular}
    \end{adjustbox}

\footnotesize{\underline{Notes:} Spatial error model estimator with two-way fixed effects. Robust standard errors are in parentheses. Time and state fixed effects are included, but not reported in the table. \\
Significance levels : *** 1\% ; ** 5\% ; * 10\%. }
\end{table}

\newpage 
\setcounter{table}{0}
\renewcommand{\thetable}{C.\arabic{table}}
\section{Analysing the effects on other frequently exposed states}
\label{appendixC}

\begin{table}[H]
\caption{Results for regressing quarterly growth on cyclone indicator}
    \centering
    \begin{adjustbox}{width= 0.48\textwidth}
        \begin{tabular}{lccc} 
        \hline \hline
              & North Carolina & Louisiana & Texas  \\
              & ARIMAX & ARIMAX  & ARIMAX  \\
              & $(4,0,0)$ & $(1,0,2)$ & $(1,0,0)$ \\
            \hline
    $\overline{Cyc}_t$  & -0.00004 & -0.00006 & -0.00007 \\
      &  (0.00003) & (0.00004) & (0.00005) \\
    $\overline{Cyc}_{t-1}$ & 0.00003 & -0.00001  & 0.00001  \\
                &  (0.00003) & (0.00004) & (0.00004) \\
    
    $\overline{Cyc}_{t-2}$ &  -0.00005 &  &   \\
                & (0.00008)  &  &  \\
    
    $\overline{Cyc}_{t-3}$ & -5.57e-06 &  &  \\
                & (0.00003) &  &   \\
    $\overline{Cyc}_{t-4}$  & 0.00002 &  &    \\
      & (0.00003) &  &  \\
      
    $\overline{Cyc}_{t-5}$   &  &  &  \\
      &  &  &  \\
    \hline
    AIC  & -1147.242 & -1222.514 & -1226.827 \\
    BIC  & -1110.743 & -1202.605 & -1210.237  \\
    
        \hline \hline 
        \end{tabular}
    \end{adjustbox}
\label{tabstates}

\footnotesize \underline{Notes:} robust standard errors in parentheses. \\
* significant at 10 \%; ** significant at 5 \%; *** significant at 1 \%

\end{table}

\begin{table}[H]
\caption{Results for regressing annual counties' growth on cyclone intensity}
    \centering   
    \begin{adjustbox}{width = 0.68\textwidth}
        \begin{tabular}{lccc}
        \hline \hline
         & North Carolina & Louisiana & Texas  \\
          & (1) & (2) & (3) \\
            \hline
             $ln(Pers.Inc.pc)_{i,t-1}$ & -0.097*** & -0.181*** & -0.221*** \\
             & (0.006) &  (0.010) & (0.006) \\
            $\overline{Cyc}_{i,t}$  & 0.00002 & 0.00004 & -0.00006 \\
              & (0.00002) & (0.00005) & (0.00008) \\
            $\overline{Cyc}_{i,t-1}$  & 0.00002** & 0.00002 & 0.00003 \\
              & (7.00e-06) & (0.00002) & (0.00003)\\
            $\overline{Cyc}_{i,t-2}$  & 0.00001** & -1.43e-06 & 0.00004* \\
              & (5.96e-06) & (0.00001) & (0.00003) \\
            $\overline{Cyc}_{i,t-3}$  & 4.98e-06 & 0.00001 & 0.00002 \\
              & (5.38e-06) & (0.00001) & (0.00002) \\
            $\rho$ & 0.118*** & 0.080*** & 0.064*** \\
              & (0.005) & (0.011) & (0.003) \\
            $\lambda$ & -0.079*** & 0.038*** & 0.064*** \\
              & (0.009) & (0.014) & (0.003) \\~\\
              
            Overall $R^2$ & 0.00 & 0.00 & 0.01 \\
            Observations & 4900 & 3136 & 9751 \\
            Number of counties/parishes & 100 & 64 & 199 \\
                        
        \hline \hline 
        \end{tabular}
    \end{adjustbox}

\footnotesize{\underline{Notes:} Spatial error model estimator with two-way fixed effects. Robust standard errors are in parentheses. Time and county fixed effects are included, but not reported in the table. \\
Significance levels : *** 1\% ; ** 5\% ; * 10\%. }

\end{table}

\end{document}